\documentclass[pra,aps,showpacs,twocolumn]{revtex4}
\usepackage{epsfig} \usepackage{graphics} \usepackage{bm}
\begin{document}
    \title{Limits to phase resolution in matter wave interferometry.}
  \author{M.~J{\"a}{\"a}skel{\"a}inen$^1$, W.~Zhang$^2$, and P.~Meystre$^1$}
\affiliation{$^1$ Optical Sciences Center, The University of Arizona, AZ 85721}
\affiliation{$^2$ Department of Physics, Tsinghua University, Beijing 100084, China}
    \date{\today}
    \begin{abstract}
We study the quantum dynamics of a two-mode Bose-Einstein
condensate in a time-dependent symmetric double-well potential
using analytical and numerical methods. The effects of internal
degrees of freedom on the visibility of interference fringes
during a stage of ballistic expansion are investigated varying
particle number, nonlinear interaction sign and strength as well as tunneling coupling. Expressions for the phase resolution are derived and the
possible enhancement due to squeezing is discussed. In particular, the role of the superfluid - Mott insulator cross-over and its analog for attractive interactions is recognized.
    \end{abstract}
    \maketitle

\section{Introduction}

Coherent atom optics offers considerable promise for applications
in a numbers of areas from precision measurements to rotation
sensors, accelerometers, and gravity
gradiometers\cite{OpticalChip,ChipReview}. One key element in any
practical device is a coherent beam splitter, and much effort has
been devoted to the realization and the understanding of these
devices, both experimentally
\cite{SplitBoulder,SplitInsbruck,SplitHannover} and theoretically
\cite{LocSplit,DoubleSplit,SplitHinds,SplitColl,SplitGarraway,SplitGirardeau,SplitBohn,SplitStickney}.

In contrast to light fields, which do not interact in a vacuum,
matter waves are subject to collisions, mostly two-body
interactions in the low density beams normally considered when
using quantum-degenerate atomic systems. At extremely low
temperatures, collisions produce a nonlinear phase shift of the
matter waves that is proportional to the atomic density, and hence
leads under normal conditions to undesirable phase noise
\cite{SearchMeystre}. This is a serious difficulty that needs to
be addressed in detail. On the other hand, two-body collisions are
also known to act as the matter-wave analog of a cubic
nonlinearity in optics. As such, they can be used to generate
nonclassical states of the Schr{\"o}dinger field. These states can
in turn be exploited to achieve phase resolution below the
standard shot-noise limit.

The use of squeezing to reduce quantum noise was suggested in \cite{Wineland1992} for Ramsey-type interferometry. For optical Mach-Zender interferometry a Heisenberg limited scheme was outlined in \cite{HollandBurnett1993}, and related schemes using dual Fock-states were suggested for Bose-condensed atoms  later. \cite{BoyerKasevich1997,DunninghamBurnett2000,DunBurnBarn2002}. 
A scheme dependent on parity measurement was suggested in \cite{GerryCampos2003}. 

A matter-wave beam splitter can be thought of as a double-well
potential with time-dependent well separation. At zero
temperature, the dominant mechanism by which atoms move from one
well to the other is quantum tunnelling. It is known that in
quantum-degenerate bosonic systems, the interplay between
tunnelling and collisions and the associated mean-field energy can
result in highly non-trivial effects \cite{Kasevich2001}. For
instance, in the case of repulsive interactions a condensate
trapped on a lattice potential can undergo a quantum phase
transition from a superfluid state, characteristic of situations
where tunneling is dominant, to a Mott insulator state,
characteristic of situations where the mean-field energy dominates
the dynamics \cite{Mott,Fisher,Zoller}. The situation is more complicated in the
case of attractive interactions, since the condensate is then
unstable unless it is small enough to be stabilized by its kinetic
energy \cite{Attractive}. Such small, stable condensates also
undergo a transition reminiscent of the superfluid-Mott insulator
transition in the sense that the ground state \cite{Cirac,Steel} changes its statistical properties at a critical value of the interaction strength.

With these processes in mind, the goal of this paper is to assess
in detail the limits in phase resolution of an atomic beam
splitter under the combined effects of collision-driven cubic
nonlinearities and quantum tunneling. The combination of these
effects renders a full quantum-mechanical description of an atomic
beam splitter highly non-trivial. Consequently, we restrict our
discussion to a two-mode analysis, using a combination of
numerical and analytical tools.

This paper is organized as follows: Section \ref{Model} discusses
our model and establishes the notation. Section \ref{Static}
presents results of a numerical analysis of the static problem
where the condensates are released from a beam splitter with fixed
well separation. Depending upon the ratio of mean field energy to
inter-well tunneling energy, the beam splitter operates either in
the superfluid or Mott insulator-like regime, with qualitative and
quantitative differences in their noise properties. For attractive
interactions, the phase noise of the beam splitter is found to be
significantly reduced at the transition between the two regimes.
These results are extended to the dynamical regime in Section
\ref{Dynamics}, which discusses in particular the departure of the
system from adiabaticity. Finally, section \ref{Conclusion} is a
summary and outlook.

\section{Model}\label{Model}

We consider in 1+1 dimensions the quantum dynamics of an ultracold
bosonic atomic beam trapped in a double-well potential $V(y,d)$
with time-dependent well separation $2d(t)$. A beam splitter using
such a time-dependent configuration of optical waveguides has been
realized by the MIT group \cite{Ketterle}, the resulting atomic
field being detected after turning off the trap and ballistic free
expansion of the atomic condensate. Assuming that the atomic
density is low enough that we can neglect three-body collisions,
the Hamiltonian of this system is
 \begin{eqnarray}
    H &=& \int_{-\infty}^{\infty}{dy {\hat{\Psi}}^{\dag}(y)\Big{[}
    -\frac{\hbar^{2}}{2M}\nabla^{2}+V(y,d)\Big{]}\hat{\Psi}(y)}\nonumber \\
    &+& g_2 \int_{-\infty}^{\infty}{dy {\hat{\Psi}}^{\dag}(y)
    {\hat{\Psi}}^{\dag}(y)\hat{\Psi}(y)\hat{\Psi}(y)},
    \label{TDSExy}
    \end{eqnarray}
where $g_2$ is the two-body coupling constant, taken to be
negative for attractive two-body interactions. In the absence of
excitations to higher spatial modes, the field operator ${\hat
\Psi}(y)$ can be expanded into two modes corresponding to
particles located around the two minima of the double-well
potential as
    \begin{equation}
        \hat{\Psi}(y) = \varphi_L(y,d)\hat{a}_L +
        \varphi_R(y,d)\hat{a}_R.
        \label{Psi_Op}
    \end{equation}
Here $\hat{a}_{L(R)}$ are bosonic annihilation operators for the
``left'' and ``right'' mode of the matter-wave field, and
$\varphi_{L(R)}$ are the corresponding spatial mode functions. For
sufficiently harmonic potentials they can be approximated by
    \begin{equation}
    \varphi(y,d)_{L(R)} = \frac{1}{(\pi\Delta y^2)^{\frac14}}
    \exp\left [-\frac{(y\pm d)^2}{4\Delta y^2}\right ],
    \label{staticmodes}
    \end{equation}
the minus sign corresponding to $\varphi_R(y,d)$ and the plus sign
to $\varphi_L(y,d)$. Within this two-mode approximation, the
Hamiltonian (\ref{TDSExy}) becomes
\begin{eqnarray}
    H(t) &=& \hbar\omega(\hat{a}_L^{\dag}\hat{a}_L +\hat{a}_R^{\dag}\hat{a}_R )
    + \frac{\Delta E(t)}{2}[\hat{a}_L^{\dag}\hat{a}_R +
    \hat{a}_R^{\dag}\hat{a}_L] \nonumber \\
    &+& g[\hat{a}_L^{\dag2}\hat{a}_L^2 + \hat{a}_R^{\dag2}\hat{a}_R^2],
    \label{Hamiltonian}
    \end{eqnarray}
where we have introduced the time-dependent tunnelling energy
\begin{widetext}
    \begin{equation}
    \Delta E(t) = \int_{-\infty}^{\infty}{dy
    \varphi_L\left (y,d(t)\right )\left (-
    \frac{\hbar^{2}}{2M}\nabla^{2}+V\left (y,d(t)\right )\right )
    \varphi_R\left (y,d(t)\right )} = \hbar\omega \exp[-d^2(t)/\Delta y^2],
    \end{equation}
\end{widetext}
and $g=g_2 d_4$ with
    \begin{equation}
    d_4 = \int_{-\infty}^{\infty}dy  \varphi^4_L(y,d(t))
    = \frac{1}{2\sqrt{\pi}\Delta y}.
\end{equation}
    Note that we have neglected cross-phase modulation,
consistently with the validity of the gaussian approximation in the
description of the mode functions of the waveguide. We remark that
this approximation only holds  for $d > \Delta x$, otherwise the
modes must be taken as time-dependent linear combinations of
the energy eigenstates, a procedure requiring numerical
diagonalization \cite{LocSplit}.

This two-mode problem is conveniently reexpressed in the Schwinger
angular momentum representation of bosonic operators
\cite{Milburn1997}. We proceed by introducing the angular momentum
operators
    \begin{equation}
    \hat{J_{z}}
    = \frac{1}{2}(\hat{a}_L^{\dag}\hat{a}_L-\hat{a}_R^{\dag}\hat{a}_R),
    \label{Jz}
    \end{equation}
    \begin{equation}
    \hat{J_{y}} =
    \frac{1}{2i}(\hat{a}_L^{\dag}\hat{a}_R-\hat{a}_R^{\dag}\hat{a}_L),
    \label{Jy}
    \end{equation}
    \begin{equation}
    \hat{J_{x}} =
    \frac{1}{2}(\hat{a}_L^{\dag}\hat{a}_R+\hat{a}_R^{\dag}\hat{a}_L),
    \label{Jx}
    \end{equation}
which can be thought of as the orthogonal components of a Bloch
vector of length $N/2$. This corresponds to mapping the
quantum state onto a distribution on the Bloch sphere.

As usual, we then express the state of the matter-wave field in
terms of eigenstates $|J, m\rangle$ of the operators ${\hat J}^2$
and ${\hat J}_z$, where
    \begin{equation} \hat{J^2}=\hat{J}_x^2+\hat{J}_y^2+\hat{J}_z^2,
    \label{J2}
    \end{equation}
    with
    \begin{eqnarray}
    {\hat J}^2|J,m\rangle &=& \hbar^2J(J+1)|J,m\rangle
    \nonumber \\
    {\hat J}_z|J,m\rangle &=& \hbar m|J,m\rangle =\frac{\hbar}{2} (n_L-n_R)|J,m\rangle,
    \end{eqnarray}
and $J= N/2$, $m=-J, -J+1, \ldots, J$.

In the angular momentum representation the Hamiltonian
(\ref{Hamiltonian}) reads
    \begin{equation}
    \label{SpinH}
    \hat{H}=f(J)+2g\hat{J}_z^2+\Delta E(t)\hat{J}_x,
    \end{equation}
where the energy $f(J)$ is a function of the total angular
momentum eigenvalue $J$. For a fixed particle number, it yields a
constant phase shift irrelevant for the problem at hand. The
ground state of the Hamiltonian (\ref{SpinH}) is expressed in
terms of the azimuthal quantum number
    \begin{equation}
    \label{GroundState}
    \vert \psi \rangle = \sum_{m=-J}^J c_m \vert
    J,m \rangle.
    \end{equation}

Each of the operators (\ref{Jz})-(\ref{Jx}) generates rotations of
this distribution around the corresponding axis. As seen from the
Hamiltonian (\ref{SpinH}), and already proposed in Ref.
\cite{Weiping} a rotation about the ${\hat x}$ axis of the Bloch
sphere can be achieved by turning on the quantum tunnelling
between the two wells for a precisely determined time. As a
result, it is possible to transform a number-squeezed state,
characterized by reduced fluctuations in ${\hat J}_z$,into a
phase-squeezed state, characterized by reduced fluctuations in
${\hat J}_y$. We exploit this feature of quantum tunnelling later
on in to achieve sub-shot noise detection in the presence of
repulsive interactions.

We mentioned that the detection of the atomic field is carried out
after the optical waveguide is rapidly switched off and the atoms
undergo a stage of ballistic expansion. The mode functions
(\ref{staticmodes}) no longer describe the spatial density of the
condensate during that stage. Rather, they must be replaced by
free gaussians that are centered around the minima of the
potential at the time of release
    \begin{eqnarray}
    \varphi_{L/R}(y,t) &=& \left (2\pi\Delta y^2(1+i\omega
    t)\right )^{-1/4} \nonumber \\
    &\times& \exp\left({-\frac{(y\pm d)^2}{4\Delta
    y^2(1+i\omega t)}}\right ).
    \end{eqnarray}
Taking the two halves of the condensate to have a relative phase
$\Theta$, the field operator for the ballistically expanding atoms
becomes then
    \begin{eqnarray}
    \hat{\Psi}(y,t,\Theta) &=&
    \hat{a}_L\varphi_L(y,t)\exp{(i\Theta/2)} \nonumber \\
    &+& \hat{a}_R\varphi_R(y,t)\exp{(-i\Theta/2)},
    \end{eqnarray}
resulting in the spatial density
\begin{widetext}
    \begin{eqnarray}
    \langle G_1(y,t,\Theta)\rangle &=&
    \langle \hat{\Psi}^{\dag}(y,t,\Theta)\hat{\Psi}(y,t,\Theta)\rangle
    ={\frac{1}{\sqrt{2\pi\Delta y^2 (1+\omega^2 t^2)}}}\exp{\Big{(}-\frac{y^2
    + d(t)^2}{2\Delta y^2 (1+\omega^2 t^2)}\Big{)}}
    \left [N\cosh \left(\frac{ y d}
    {\Delta y^2(1+\omega^2 t^2)}\right) \right . \nonumber \\
    &+& \left . 2\langle\hat{J}_x\rangle\cos\left(
    \frac{ y d}{\Delta y^2 (1+\omega^2 t^2)}\omega t+\Theta\right )
    -  2\langle\hat{J}_y\rangle\sin\Big{(}\frac{ y d}
    {\Delta y^2(1+\omega^2 t^2)}\omega t+\Theta\Big{)} \right ].
    \label{G1}
    \end{eqnarray}
\end{widetext}
The atomic density at any point in space and time is given by an
incoherent contribution that is independent of both the relative
phase and the internal dynamics of the two-mode condensate, as
well as a coherent contribution.

Since the Hamiltonian ${\hat H}$ is invariant with respect to the
exchange $L \leftrightarrow R$, it is easily seen that for states
symmetric with respect to the interchange of the two modes,
$\langle {\hat J}_y\rangle=\langle {\hat J}_z\rangle=0$ for all
times, although $\langle {\hat J}_y^2\rangle \neq 0$ as we discuss
shortly. The coherent contribution to $\langle
G_1(y,t,\Theta)\rangle$ is therefore proportional to the
expectation value of $\langle \hat{J}_x \rangle $. In terms of the
angular momentum picture, it can be interpreted as the
polarization of the distribution. The visibility of the
interference fringes of the ballistically expanding two-mode
condensate,
    \begin{equation}
    \label{visibility}
    V=\frac{\langle
    G_1(0,t,0)\rangle-\langle G_1(0,t,\pi)\rangle}{\langle
    G_1(0,t,0)\rangle+\langle G_1(0,t,\pi)\rangle}=
    \frac{\vert\langle\hat{J}_x\rangle\vert}{N/2},
    \end{equation}
depends only on that expectation value, the associated
fluctuations being given by
    \begin{eqnarray}
    \Delta G_1&\equiv&\sqrt{\langle\Delta
    G_1^2(y,t,\Theta)\rangle} \nonumber \\
    &=& \sqrt{\langle G_1^2(y,t,\Theta)\rangle
    - \langle G_1(y,t,\Theta)\rangle^2}. \label{dG1}
    \end{eqnarray}
The phase resolution of atom-interferometric experiments is
limited by the requirement that the change in local density
resulting from an imprinted global phase change must be larger
than the intrinsic fluctuations of the first order correlation
function
    \begin{equation}
    \Delta G_1 \ge \left |\frac{\partial
    \langle G_1\rangle}{\partial \Theta}\right | \Delta\Theta.
    \end{equation}
This gives as an estimate for the phase resolution resulting from
number fluctuations
    \begin{equation}
    \Delta \Theta = \frac{\sqrt{\langle\Delta
    G_1^2\rangle}}{\left |\partial \langle
    G_1\rangle/\partial \Theta \right |}.
    \end{equation}

Applying this criterion to the problem at hand, one finds that the
phase resolution for the two-mode condensate is given by
    \begin{equation}
    \label{PhSq}
    \Delta \Theta^2 =
    \frac{\langle \hat{J}_y^2\rangle}{\langle
    \hat{J_x}\rangle^2}+\frac{\langle\hat{N}^2\rangle-
    \langle\hat{N}\rangle^2}{4 \langle
    \hat{J_x}\rangle^2}.
    \end{equation}

The second term in Eq.(\ref{PhSq}) is equal to zero in an
individual experimental event, but gives rise to a contribution of
order unity if averaged over a classical (Poissonian) distribution
of particle numbers in repeated experiments. The number of
particles must therefore be determined with a high precision if
one wishes to benefit from any enhancement of interferometric
resolution due to phase squeezing. Setting this second term to
zero for now, we recover the result derived by Kitagawa and Ueda
in their seminal work on spin squeezing \cite{Ueda1993}. Squeezing
along the $\hat{y}$-direction corresponds to increased
correlations of the phases between the two wells, thus producing a
better defined relative phase, a desirable feature in
interferometric applications. The squeezing parameter, which we
calculate relative to the standard quantum limit
    \begin{equation}
    \xi_y  = \frac{\Delta \Theta}{\Delta \Theta_{SQL}} =
    \frac{\sqrt{N\langle \hat{J}_y^2\rangle}}{\langle \hat{J_x}\rangle},
    \label{Squeeze}
    \end{equation}
    although not necessarily representing the maximal achievable
squeezing of the variances of the angular momentum distribution,
was shown by these authors to represent the parameter of interest
from the point of view of possible interferometric applications.
Hence it is the major focus of the present study.
    \begin{figure}[t]
    \begin{center}
    \epsfig{file=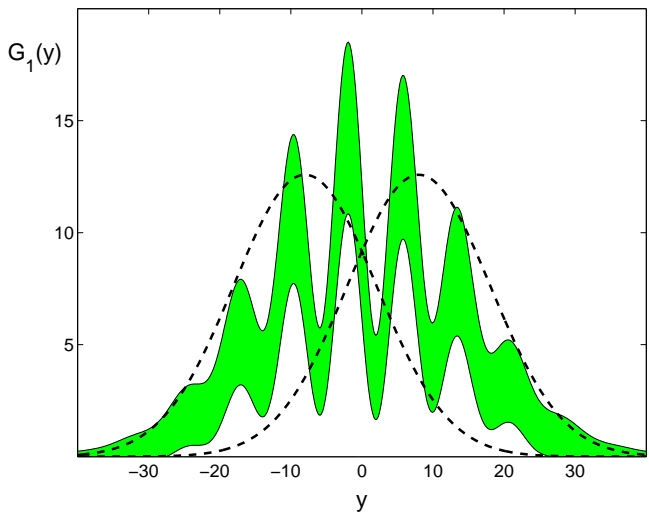,width=8.6cm}
    \caption{\label{fig:G1}Particle density bounded by
    density noise $\langle G_1\rangle \pm\Delta G_1$
    after ballistic expansion of a two-mode condensate with $N = 100$ in the
    'superfluid' regime where the visibility is high. The dashed lines
    show the density of the mode functions. Transverse distance is
    measured in units of the oscillator width $\Delta y$ and the
    initial condensates were centered around $d = \pm8\Delta y$.}
    \end{center}
    \end{figure}

\section{Results}
\label{Results}

This section discusses the main results of our numerical study of
the two-mode beam splitter. The analysis is based on the
dimensionless parameter
    \begin{equation}
    G = \frac{2 g N}{\Delta E},
    \label{G_definition}
    \end{equation}
the ratio of mean-field energy to tunnelling energy, which completely
determines the spectrum of the Hamiltonian.

Note however that the two-mode approximation implicitly
assumes a stable condensate, a property that holds only for low
enough densities in the case of attractive interactions \cite{Attractive}.

We consider first the simple case of a static double-well system
where the pconfining potential is suddenly turned-off, resulting into the formation of an interference pattern after a period of ballistic expansion. The
next subsection discusses the results of a full dynamical study.

\subsection{Static double-well potential}\label{Static}

The matter-wave interference pattern is shown in Fig.
{\ref{fig:G1} in the superfluid regime, further illustrating the
excellent contrast in that case. In addition to the fringe
contrast, it is necessary to consider the quantum fluctuations of
the interference pattern, since they lead to the fundamental limit
in phase resolution of interferometric measurements. Figure
\ref{fig:G1} illustrates these fluctuations by attaching to the
intensity $\langle G_1 \rangle$ a width given by twice its
variance $\Delta G_1 = (\langle G_1^2 \rangle - \langle
G_1\rangle^2)^{1/2}$. We have seen in Eq. (\ref{PhSq}) that these
fluctuations result in a phase resolution
    \begin{equation}
    \Delta \Theta^2 =
    \frac{\langle \hat{J}_y^2\rangle}{\langle
    \hat{J_x}\rangle^2},
    \end{equation}
where we have neglected the shot-to-shot number fluctuations for
simplicity. The density noise is here proportional to
$\sqrt{\langle G_1\rangle}$, allowing at best for interferometry
at the standard quantum limit.

Figures \ref{fig:AttrGroundStates} and \ref{fig:RepGroundStates} shows the ground state distributions in terms of both the probability amplitudes $|c_m|$, and of the
amplitudes $|c_{\theta_m}|$ of the ground state expressed on a
basis of so-called relative phase states,
    \begin{equation}
    |\psi\rangle = \sum_m c_{\theta_m} |\theta_m\rangle,
    \label{state theta}
    \end{equation}
    where \cite{LuisSS}
    \begin{equation}
    \label{PhaseDistr}
    |\theta_m\rangle = \frac{1}{\sqrt{2J+1}}\sum_{m^{\prime} =
    -J}^J\exp(im^{\prime}\theta_m)\vert J,m^{\prime}\rangle
    \end{equation}
and the discrete relative phases are given by
    \begin{equation}
    \theta_m = \theta_0 +\frac{2\pi m}{2J+1},
    \end{equation}
for an arbitrary reference phase $\theta_0$ chosen here to be
zero.

For weak interactions, i.e. small $\vert G\vert$, the system is in
a state reminiscent of a coherent state, with a relatively well
defined phase with fluctuations consistent with the standard
quantum limit. For large attractive interactions the ground state
approaches a double peaked distribution, corresponding to a
macroscopic superposition state as described in
\cite{Cirac,Steel}. For $G \ll -1$, the ground state can thus  be
approximated by
    \begin{equation}
    \label{PosCat}
    \vert\psi\rangle =
    \frac{1}{\sqrt{2}}\big{[}\vert J,J\rangle+\vert
    J,-J\rangle\big{]}.
    \end{equation}
This gives for the phase distribution
    \begin{equation} \label{PhaseDistrAttr}
    \vert\psi\rangle =
    \frac{1}{\sqrt{J+1/2}}\sum_{m=-J}^J\cos(J\theta_m)\vert\theta_m\rangle.
    \end{equation}
The transition between the superfluid and superposition regimes
takes place just below $G = -1$, where the relative phase becomes
well defined due to squeezing.

For large repulsive interactions the ground state goes through the
Mott-insulator transition in a continuous manner. Here the number
distribution narrows and the phase distribution widens until it
becomes essentially flat, indicating a completely random phase
when averaged over an ensemble.
    
    \begin{figure}[t]
    \begin{center}
    \epsfig{file=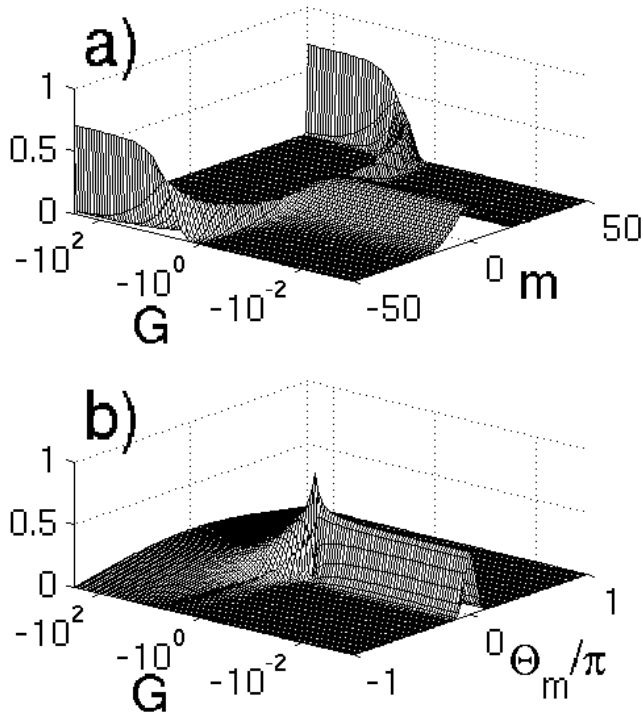,width=8.6cm}
   \caption{\label{fig:AttrGroundStates} Ground-state distributions as function of the dimensionless parameter $G$ for attractive interaction. In a) magnitudes of ground state components $\vert c_m(G)\vert$ are shown for attractive interaction, and in b) the distribution of relative phase components $\vert c_{\theta_m}(G)\vert$ in Eq.(\ref{PhaseDistr}).}
 \end{center}
\end{figure}
\begin{figure}[t]
    \begin{center}
    \epsfig{file=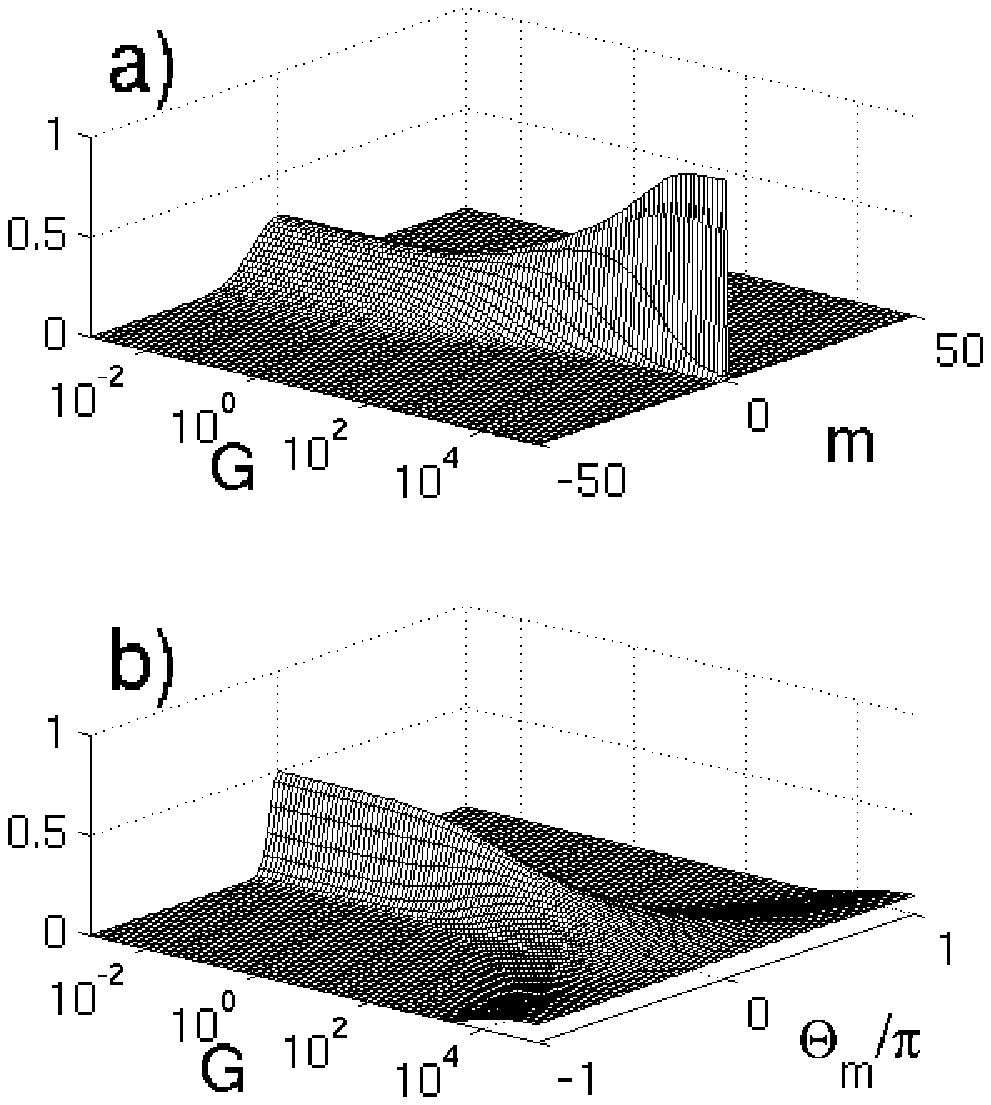,width=8.6cm}
   \caption{\label{fig:RepGroundStates} Ground-state distributions as function of the dimensionless parameter $G$ for repulsive interaction. In a) magnitudes of ground state components $\vert c_m(G)\vert$ are shown for attractive interaction, and in b) the distribution of relative phase components $\vert c_{\theta_m}(G)\vert$ in Eq.(\ref{PhaseDistr}).}
 \end{center}
\end{figure}

  \begin{figure}[t]
    \begin{center}
    \epsfig{file=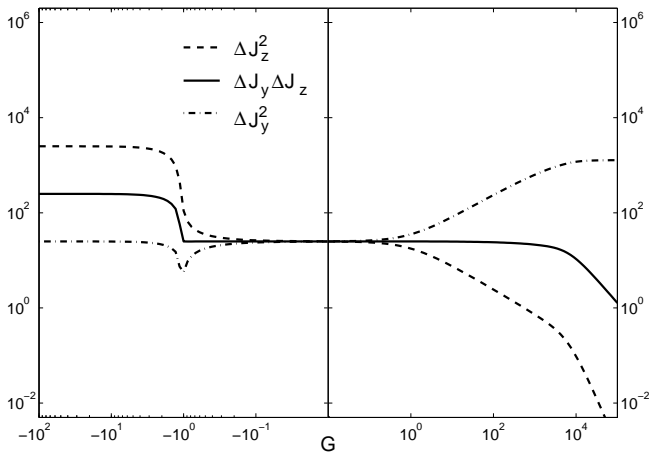,width=8.6cm}
    \caption{\label{fig:UncRelPlot}The uncertainties $\Delta\hat{J}_y^2$, $\Delta\hat{J}_z^2$ and the product $\Delta\hat{J}_y\Delta\hat{J}_z$ as functions of $G$ for $N = 100$. The phase uncertainty $\Delta\hat{J}_y^2$ goes through a global minimum at $G = -1$, the border between attractive superfluid and superposition states, and increases in the repulsive regime until it saturates to a constant value as the system passes through the Mott-insulator transition.}
    \end{center}
\end{figure}

Figure \ref{fig:UncRelPlot} shows the uncertainties in the
relative number $\Delta\hat{J}_z^2$, and the relative phase
$\Delta\hat{J}_y^2$, together with their product
$\Delta\hat{J}_y\Delta\hat{J}_z$ for $N = 100$ atoms. The
uncertainty product is essentially constant in the region
    \begin{equation} \label{SF_Regime}
    -1 \ll G \ll \frac{1}{2}N^2,
    \end{equation}
where the upper limit was found from the simulations. In this
interval the ground state resembles a coherent state, making this
the superfluid regime. For large attractive interactions the
uncertainty product increases at the superfluid-superposition
transition and saturates for strong interactions. As the
uncertainty in relative phase does not increase when compared to
its value in the superfluid regime, the increase in the
uncertainty product can be attributed solely to the increase in
$\Delta\hat{J}_z$. The fact that the ground state in the
superposition regime is double peaked, as can be seen in Fig.
\ref{fig:AttrGroundStates} and thus no longer a minimum uncertainty
state, gives using Eq.(\ref{PosCat}) for the uncertainty in number
difference \begin{equation} \langle\hat{J}_z^2\rangle = J^2 =
\frac{N^2}{4}, \end{equation} in agreement with the numerical
results, see for instance Fig. \ref{fig:UncRelPlot}. On the
Bloch-sphere the distribution is concentrated around the two
regions $m = \pm J$, this bimodal character giving giving rise to
a large spread in $\Delta\hat{J}_z$.

\begin{figure}[t]
    \begin{center}
    \epsfig{file=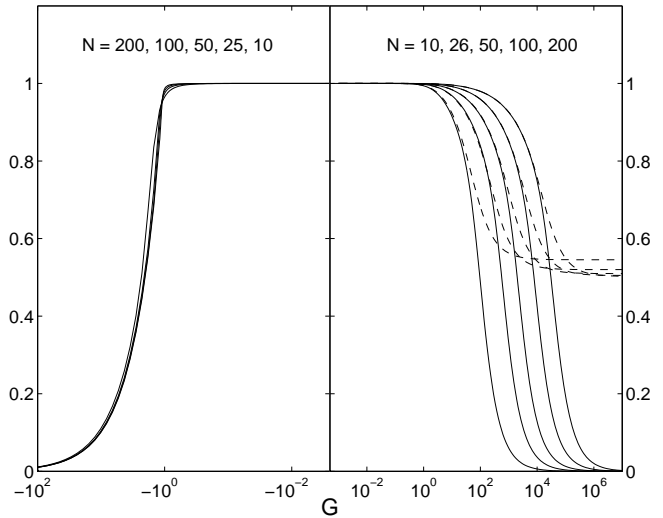,width=8.6cm}
    \caption{\label{fig:Vis} Visibility of interference fringes for a
ballistically expanding two-mode BEC as a function of $G$, the
ratio between the mean-field and tunneling energies, for particle
numbers ranging from $10$ to $200$. For attractive interaction the
visibility decreases abruptly for $G < -1$. In the case of
repulsive interaction the behavior is qualitatively and
quantitatively different for even (solid) and odd (dashed) atom
numbers, as explained in the text.}
    \end{center}
\end{figure}

Figure \ref{fig:Vis} shows the visibility $V$ of the fringes, Eq.
(\ref{visibility}), after the free expansion of the matter waves
following the switching off of a static double-well potential. In
case tunnelling dominates the dynamics of the system, we observe
high-contrast interference fringes, but there is an abrupt
transition, with contrast decreasing as $\vert G\vert^{-1}$ as
soon as $G < -1$. The underlying physics governing the decrease of
contrast is essentially the same as in the superfluid-Mott
insulator transition predicted \cite{Fisher,Zoller} and observed
in optical lattices \cite{Mott}. If tunnelling dominates, the
state of the two-mode system is essentially a superfluid with a
well-established phase relationship between the two modes of the
beam splitter, resulting in high-contrast interferences. In the
superposition and Mott regimes, by contrast, the two wells are
isolated from each other, with no phase relationship between them.

This behavior of the fringe visibility can be further understood
from the Heisenberg uncertainty relation
    \begin{equation}
    \label{UncRelation}
    \Delta\hat{J}_y\Delta\hat{J}_z =
    \frac{1}{2}\vert\langle[\hat{J}_y,\hat{J}_z]\rangle\vert =
    \frac{1}{2}\vert\langle\hat{J}_x\rangle\vert,
    \end{equation}
which shows that the uncertainty product
$\Delta\hat{J}_y\Delta\hat{J}_z$ is proportional to the
polarization $\langle\hat{J_x}\rangle$ and thus the visibility
(\ref{visibility}). This shows that the regime of high visibility
coincides with the one where uncertainty product is constant. This
is, as expected, the superfluid regime. The polarization measures
the coherence, here the degree to which adjacent wave function
components $c_m$ are populated.
In the Mott-insulator and superposition regimes, the wave function
becomes highly peaked around the center (minimal $\vert m\vert$)
and boundary (maximal $\vert m\vert$) which thus limits the
coherence.

In addition to illustrating that the visibility decreases outside
the superfluid regime given by Eq. (\ref{SF_Regime}),
Fig.\ref{fig:Vis} also shows that for $G > 0$ and even particle
number, the asymptotic value for the visibility is $V=\frac{1}{2}$
rather than zero for $N$ odd. This difference in asymptotic
behaviors can be explained as follows. For repulsive interactions
and far into the Mott insulator regime, the ground state is
approximately be given by a Fock state with equal populations in
both wells. In the case of odd atom number, the additional atom
can be in either of the wells, so that
    \begin{equation}
    \label{OddState}
    \vert\psi\rangle = \frac{1}{\sqrt{2}}\Big{[}\vert
    J,\frac{1}{2}\rangle+\vert J,-\frac{1}{2}\rangle\Big{]}.
    \end{equation}
This gives for the polarization
    \begin{equation}
    \label{OddVis}
    \langle \hat{J}_x\rangle =
    \frac{1}{2}\big{[}J+\frac{1}{2}\big{]} \approx \frac{J}{2},
    \end{equation}
and hence $V = 1/2$ as asymptotic value.

\begin{figure}[t]
    \begin{center}
    \epsfig{file=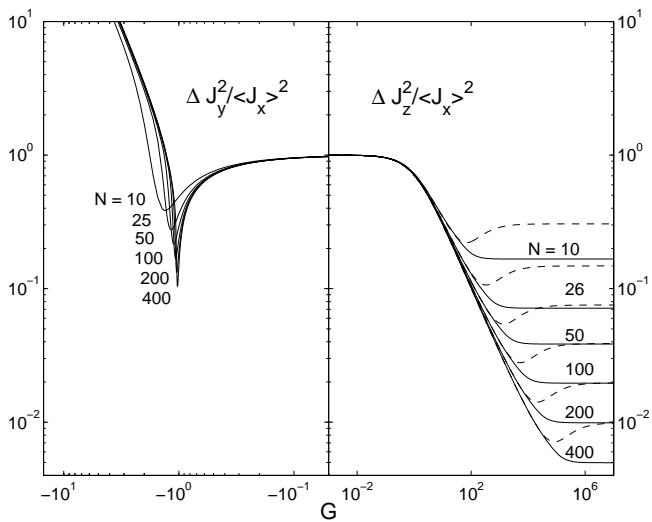,width=8.6cm}
\caption{\label{fig:QPT} Squeezing of the phase variance relative
to the standard quantum limit for the two-mode ground state. For
attractive interaction the phase squeezing is minimal just below
$G = -1$. The crossover from Poissonian fluctuations in the
superfluid regime $G < -1$ to the superposition state $G > -1$
becomes sharper with increased number of particles. For repulsive
interaction there is a smooth transition between the superfluid
regime of low G-values to the Mott-insulator state. In the
asymptotic limit $G\rightarrow\infty$, the squeezing approaches a
plateau, which is different for odd and even atom numbers.}
    \end{center}
\end{figure}

For the case of even particle number, in contrast, the asymptotic
ground state for $G\rightarrow\infty$ is the Fock state $\vert m =
0\rangle$ with $N/2$ atoms in each well. For large but finite
values of $G$ the state can be approximated by
    \begin{equation} \label{EvenState}
    \vert\psi\rangle = \sqrt{1-2\varepsilon}\vert
    J,0\rangle-\sqrt{\varepsilon}\Big{[}\vert J,1\rangle+\vert
    J,-1\rangle\Big{]},
    \end{equation}
where epsilon is a small number. This gives for the polarization
    \begin{equation}
    \label{EvenVis}
    \langle \hat{J}_x\rangle =
    2\sqrt{\varepsilon}\sqrt{J(J+1)} \approx 2\sqrt{\varepsilon}J,
    \end{equation}
and an asymptotic value of the visibility $V \rightarrow 0 $.

Let us now turn to the phase resolution $\Delta \Theta^2$ of the
beam splitter. The left-hand side of Fig. \ref{fig:QPT} shows the
variance $\Delta\Theta^2/\Delta\Theta^2_{\rm SQL}$ after free
ballistic expansion from the two-mode ground state of the beam
splitter, plotted as a function of the ratio $G$ between the mean
field energy and tunnelling energy. The phase resolution exhibits
a sharp minimum just below $G = -1$. At this point, we find
numerically that the energy gap between the ground state and the
first excited state of the double-well goes to zero within our
numerical accuracy. This indicates the presence of a `quantum
phase transition' --- or more precisely cross-over --- a property
also inferred from Fig. \ref{fig:AttrGroundStates} where the quantum
statistical properties of the ground state obviously change at
the point $G = -1$. This transition is associated with the onset
of increased number fluctuations between the populations of the
two modes, and correspondingly to suppressed phase fluctuations.
Hence, the phase resolution of the beam splitter is maximized at
that point. As expected, the crossover from Poissonian
fluctuations in the superfluid regime for $G > -1$ to a
superposition state for $G < -1$ becomes sharper with increased
particle number. For attractive interaction there is thus a
squeezed state at the boundary between the superfluid and
superposition regimes. For small atom numbers and attractive
interaction, the minimum of $\Delta\Theta^2$ occurs to the right
of the superfluid-superposition transition. From this discussion,
it might appear favorable to operate the beam splitter in that
regime. This is however misleading, as we must also take into
account the fact that the fringe visibility rapidly decreases in
that regime, as shown in Fig. \ref{fig:Vis}. The situation
improves rapidly for large $N$, though. The sharpness of the
minimum, however, makes this state challenging to create
experimentally in a controlled manner. In addition, the maximum
achievable squeezing is limited due to the size limits imposed by
the metastability of attractive condensates.

\begin{figure}[t]
    \begin{center}
    \epsfig{file=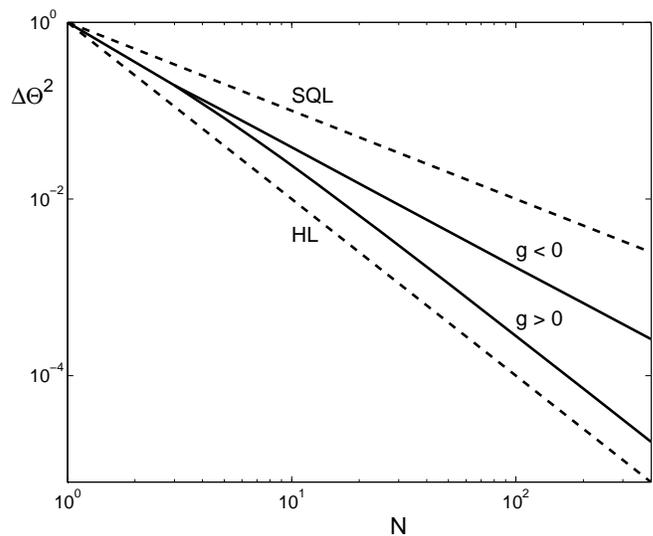,width=8.6cm}
\caption{\label{fig:Scaling} Phase variance in the ground state of
the double-well system at the ``critical point'' $G = -1$ for
attractive condensates, and at $G = \frac{N^2}{2}$ for repulsive
condensates, as functions of particle number. The dashed lines
show the standard quantum limit (SQL) and Heisenberg limit (HL)
for comparison.}
 \end{center}
 \end{figure}

The situation is slightly more subtle in the case of repulsive
interactions. From Fig. 2, it is quite clear that as the system
moves into the Mott regime, the phase uncertainty increases and
the number fluctuations of the ground state become more strongly
squeezed. This is also evidenced in Fig. 3, which shows the
monotonic increase in $\Delta {\hat J}_y^2$ for increasing $G$.
Hence, it would appear that the phase resolution becomes
increasingly worse in this regime. This difficulty can however be
eliminated by using quantum tunnelling to turn number squeezing
into phase squeezing, as discussed in section II. As a result, the
minimal obtainable phase fluctuations for $G > 0$ are given by
    \begin{equation}\label{MinRep}
    \Delta\Theta^2 =
    \frac{\Delta\hat{J}_z^2}{\langle\hat{J}_x\rangle^2}
    \end{equation}
since $\Delta\hat{J}_z = \langle\hat{J}_z^2\rangle$ is the minimal
uncertainty of ${\hat J}_z$ in deep the Mott regime. The
right-hand side of Fig. \ref{fig:QPT} shows that the phase
resolution achieved by this technique in the case of repulsive
interactions. Just as for the visibility, we find here different
asymptotic behavior for odd and even particle numbers with a
factor of two in difference for large interactions. Using
Eq.(\ref{OddState}) we find for the uncertainty in number
difference
    \begin{equation} \label{Jz2Odd}
    \langle\hat{J}_z^2\rangle = \frac{1}{4},
    \end{equation}
which together with Eq.(\ref{OddVis}) for the polarization gives
    \begin{equation} \label{OddPhDisp}
    \frac{\langle\hat{J_z}^2\rangle}{\langle\hat{J_x}\rangle^2}
    \approx \frac{1}{J^2} = \frac{4}{N^2},
    \end{equation}
in agreement with the asymptotic values for odd particle numbers, shown dashed in Fig. \ref{fig:QPT}. For
even numbers, Eq.(\ref{EvenState}) yields
    \begin{equation}
    \label{EvenJz2}
    \langle{\hat{J_z}^2}\rangle \approx
    2\varepsilon,
    \end{equation}
which together with Eq.(\ref{EvenVis}) gives
    \begin{equation}
    \label{EvenPhDisp}
    \frac{\langle\hat{J_z}^2\rangle}{\langle\hat{J_x}\rangle^2}
    \approx \frac{1}{2J^2} = \frac{2}{N^2},
    \end{equation}
in agreement with Fig.\ref{fig:QPT}. The behavior of both phase
resolution and visibility for odd versus even particle numbers can
thus be understood from the form of the ground state in the limit
of large repulsive interactions.

Strictly speaking, $\Delta\hat{J}_y$ is the phase uncertainty only
when the distribution spans an area narrower than the diameter of
the Bloch sphere. For larger values the spread $\Delta\hat{J}_y$
behaves differently, as it is the projection onto the plane
spanned by $\hat{J}_y$ and $\hat{J}_z$, and not the distance along
the equator. For the interferometric setup considered here the
phase uncertainty is in general not equal to the phase resolution
given by Eq. (\ref{PhSq}), as the latter will grow when the
visibility goes down.
    
\begin{figure}[t]
    \begin{center}
    \epsfig{file=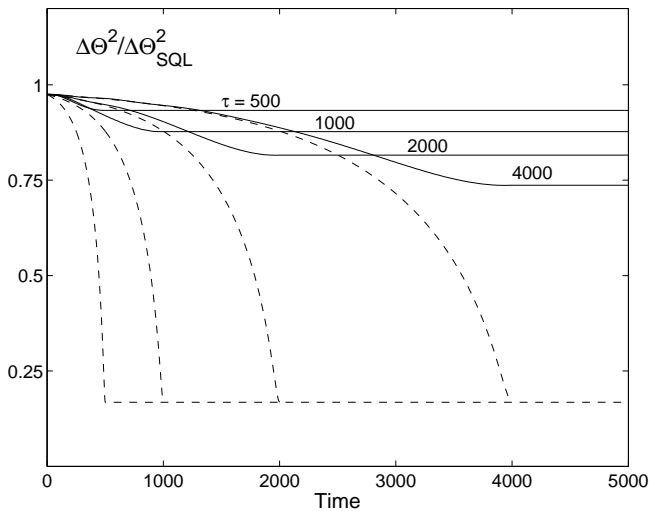,width=8.6cm}
\caption{\label{fig:SqueezeDynAttr} Phase squeezing for $g < 0$ when $\Delta E(t)$ is taken to vary according to Eq.(\ref{d_time_dep}),  for several values of the
$\tau$. The dashed lines show squeezing of
the instantaneous ground state for comparison.}
    \end{center}
\end{figure}

\begin{figure}[t]
    \begin{center}
    \epsfig{file=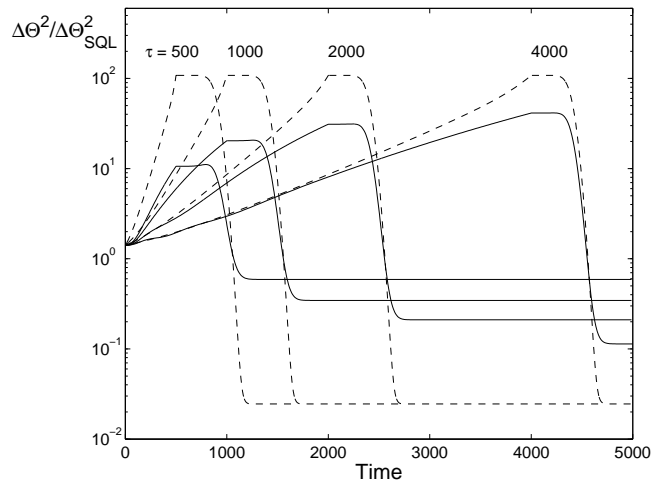,width=8.6cm}
\caption{\label{fig:SqueezeDynRep} Squeezing of the relative phase
$\Delta \Theta$ for a double well split in time according to
equation Eq.(\ref{d_time_dep}).}
 \end{center}
\end{figure}

We have seen in Section II that the phase sensitivity of the
system is closely related to spin squeezing. The squeezing of both
optical and matter waves is of considerable interest in
interferometry, as it offers the potential to beat the standard
quantum limit of detection, and possibly replace it by the
so-called Heisenberg limit scaling as $1/N$. Fig. \ref{fig:Scaling} shows the phase
variation of the ground state at both the ``critical point'' $G =
-1$ for attractive interaction, and at $G = \frac{1}{2}N^2$ for
repulsive interaction as functions of particle number $N$, the
dashed lines giving the standard quantum limit (SQL) scaling as
$1/\sqrt{N}$ and Heisenberg limit (HL) for comparison. The G-value in the repulsive regime was chosen to produce maximal squeezing for odd particle numbers, slightly worse than for even numbers. Since the particle number can not be controlled down to single units, this produces a conservative limit for the achievable squeezing. 

The combination of the mean-field interaction and
tunneling clearly leads to the squeezing of the matter-wave field
and a sensitivity significantly improved from the standard quantum
limit under the conditions mentioned above. We find numerically that for attractive interaction $\Delta\Theta^2$ scales as $N^{-1.38}$, slightly above the
asymptotic limit of $N^{-4/3}$ derived by Kitagawa and Ueda
\cite{Ueda1993}. The difference is expected to decrease with
increased particle number. For repulsive interaction we find
$\Delta\Theta^2 \propto N^{-2}$, thus corresponding to Heisenberg
limited phase resolution. This comes at the price of larger
interactions and also necessitates an additional rotation of the Bloch sphere.

\subsection{Dynamics}\label{Dynamics}
The previous section investigated the ground state properties of
the static double-well system. Here we discuss ways to create states with the desired properties starting from a condensate in the ground state with
a small interwell separation. The adiabatic theorem of quantum
mechanics states that a system governed by a time dependent
Hamiltonian and initially prepared in an eigenstate will remain in
the instantaneous eigenstate given that the Hamiltonian changes
sufficiently slowly. The aim here is evolve the system
adiabatically \cite{Javanainen} into states of maximal squeezing and then freeze the dynamics.

Controlling the magnitude and sign of two-body interaction is
readily achievable using Feshbach resonances. The rate at which
the magnetic field can be swept across a resonance is much higher
than typical trap frequencies and tunnelling rates, allowing for a
practically instantaneous switching of the nonlinear interaction.
Decreasing the tunnelling rate can be achieved by rapidly
separating the two wells as the point of minimal phase
fluctuations is reached.

We model the splitting of the symmetric double-well potential
using a tunnelling potential $\Delta E(t)$ that decreases exponentially
in time
    \begin{equation}
    \Delta E(t) = \hbar \omega \exp{ \Big{(} -\frac{d_{\rm min}^2}{\Delta y^2}-\Gamma t}\Big{)},
    \label{d_time_dep}
    \end{equation}
where $d_{\rm min}$ is the minimal separation between the wells,
and the constant $\Gamma$ is a measure of how fast the wells are
separated. The separation is assumed to occur over a finite time
$\tau$ and to reaching a value $d(\tau)$ such that  $G = -1$ for
attractive interaction, and $G = \frac{1}{2}N^2$ for repulsive
interaction. The constant $\Gamma$ is adjusted accordingly. As
the time $t = \tau$ is reached, the two-body interaction is
rapidly taken to zero to freeze the dynamics.

The evolution after $t = \tau$ is linear and solely governed by
tunnelling. Note that for attractive interactions, the residual
rotation of the distribution resulting from quantum tunnelling has
to be kept small in order to avoid transforming the phase
squeezing into number squeezing,
    \begin{equation}\label{NoPulse}
    \int_{\tau}^{\infty}\Delta E(t)dt
    \ll \hbar\pi,
    \end{equation}
This can be achieved by either taking $\Delta E(\tau) \ll 1$ at
the point where $G = -1$, or separating the wells rapidly after $t
= \tau$ to make the left-hand side of Eq. (\ref{NoPulse})
negligible.

The squeezing in the phase variation for attractive interaction is
shown versus time in Fig. \ref{fig:SqueezeDynAttr} for several
values of the parameter $\tau$. The instantaneous squeezing of the
true ground state is shown dashed for comparison. The evolution of
the system is initially adiabatic, but becomes diabatic for larger
separations making the system freeze out in a state less squeezed
than the desired ground state. Longer evolution times make the
system more adiabatic thus achieving squeezing closer to, but
still far from the optimal values. For all cases shown in Fig.
\ref{fig:SqueezeDynAttr} it is however apparent that the dynamics
is only partially adiabatic and that the minimal phase variation
is severely limited by violation of  the requirement that the
dynamics be adiabatic.

Figure \ref{fig:SqueezeDynRep} shows the squeezing dynamics of a
condensate with repulsive interaction when the tunnelling energy
is decreased exponentially in time according to
Eq.(\ref{d_time_dep}). As the value $G = \frac{1}{2}N^2$ is
reached, the two-body interaction is rapidly tuned to zero. After
that a pulse of tunnelling interaction, which corresponds to
bringing the two wells closer again, is applied in such a way that
    \begin{equation}\label{PiPulse}
    \int_{\tau}^{\infty}\Delta E(t)dt
    = \frac{\hbar\pi}{2},
    \end{equation}
thereby rotating the Bloch-sphere distribution by an angle $\pi/2$
around the $\hat{J}_x$-axis and thus transforming the number
squeezing into phase squeezing without otherwise changing the
distributions.

Initially the phase variance is seen to increase as the state
becomes number-squeezed for increasing values of $G(t)$ until the
time $t = \tau$ is reached when $g \rightarrow 0$, effectively
freezing the evolution until the tunnelling pulse (\ref{PiPulse})
is applied. For all values of $\tau$, which are indicated in
Fig.\ref{fig:SqueezeDynRep}, the same tunnelling pulse was used.
The dynamics is seen in Fig. \ref{fig:SqueezeDynRep} to become
more and more adiabatic as the value of $\tau$ is increased, just
as in the case of attractive interaction. Here, however, the
maximal squeezing achieved is larger than for the attractive case.
This is not obvious at first, even though the repulsive ground
state is more squeezed, as the system has to be evolved to values
of stronger interaction than in the attractive case.

\section{Conclusion and outlook}\label{Conclusion}

In this paper we have investigated the full quantum dynamics of a
condensate in a symmetrically split double well. Expressions for
the visibility and phase resolution during a ballistic expansion
stage were given and investigated numerically. The possibility of
creating phase squeezed ground states by adiabatic splitting was
demonstrated, but is limited by the time scales involved. The
increased phase sensitivity for ground states of attractive
condensates, which are known to be stable only for particle
numbers \cite{Attractive} up to around $N \approx 10^3$, was found
to require long splitting times to achieve adiabatic evolution.
For repulsive condensates where the densities are limited only by
the requirement that the two-mode model be applicable, a scheme
with scaling at the Heisenberg limit was outlined and tested in
dynamics simulations. The original Heisenberg-limited scheme suggested in \cite{HollandBurnett1993} for optical MZ interferometers and later applied in various forms to the case of atomic condensates \cite{BoyerKasevich1997,DunninghamBurnett2000,DunBurnBarn2002}, used dual Fock-states. i.e. the state was of the form given by Eq.(\ref{EvenState}). In the present context such states are unsuitable due to their low visibility. The problems can be avoided by not taking the system deep into the Mott-insulator regime where the ground state is the dual Fock state, but rather to the threshold where $G = N^2/2$, and the visibility still is high. 

It is well known that the introduction of a linear
potential, for instance due to gravity, changes the localization
properties of the double-well eigenfunctions. The corresponding
effects in the many-body regime are presently explored
experimentally \cite{Kasevich2} and will also be the subject of
future theoretical investigations. \acknowledgments

This work is supported in part by the US Office of Naval Research,
by the National Science Foundation, by the US Army Research
Office, by the National Aeronautics and Space Administration, and
by the Joint Services Optics Program.


\begin{thebibliography}{99}
\bibitem{OpticalChip} G. Birkl, F. B. J. Buchkremer, R. Dumke, W.
Ertmer, Opt. Comm. \textbf{191}, 67 (2001).
\bibitem{ChipReview} J. Reichel,  Appl. Phys. B \textbf{75}, 469
(2002).
\bibitem{SplitBoulder} D. M{\"u}ller, E. A. Cornell, M.
Prevedelli, P. D. D. Schwindt, A. Zozulya, and D. Z. Anderson,
Opt. Lett. \textbf{25}, 1392 (2000).
\bibitem{SplitInsbruck} D. Cassettari, B. Hessmo, R. Folman, T.
Maier, J. Schmiedmayer, Phys. Rev. Lett. \textbf{85}, 5483 (2000).
\bibitem{SplitHannover} R. Dumke, T. M{\"u}ther, M. Volk, W.
Ertmer, and G. Birkl, Phys. Rev. Lett. \textbf{89}, 220402 (2002).
\bibitem{LocSplit} M. J{\"a}{\"a}skel{\"a}inen, and S. Stenholm,
Phys. Rev. A \textbf{68}, 033607 (2003).
\bibitem{DoubleSplit} L. A. Collins, L Pezz{\'e}, A. Smerzi, G. P.
Berman, and A. R. Bishop, {\tt quant-th/0404149}.
\bibitem{SplitHinds} E. A. Hinds, C. J. Vale, and M. G. Boshier,
Phys. Rev. Lett. \textbf{86}, 1462 (2001).
\bibitem{SplitColl} E. Andersson, T. Calarco, R. Folman, M.
Andersson, B. Hessmo, and J. Schmiedmayer, Phys. Rev. Lett.
\textbf{88}, 100401 (2002).
\bibitem{SplitGarraway} O. Zobay, and B. M. Garraway, Opt. Comm.
\textbf{178}, 93 (2000).
\bibitem{SplitGirardeau} M. D. Girardeau, K. K. Das, and, E. M.
Wright, Phys. Rev. A \textbf{66}, 023604 (2002).
\bibitem{SplitBohn} D. C. E. Bortolotti, and, J. L. Bohn, Phys.
Rev. A \textbf{69}, 033607 (2004).
\bibitem{SplitStickney} J. A. Stickney, and A. A. Zozulya, Phys.
Rev. A \textbf{68}, 013611 (2003).
\bibitem{SearchMeystre} C. P. Search, and P. Meystre, Phys. Rev. A
\textbf{67}, 061601 (2003).
\bibitem{Wineland1992} D. J. Wineland, J. J. Bolinger, W. M. Itano, F. L. Moore,  and D. J. Heinzen, Phys. Rev A \textbf{46}, R6797 (1992).
\bibitem{HollandBurnett1993} M. J. Holland and K. Burnett, Phys. Rev. Lett. \textbf{71}, 1355 (1993).
\bibitem{BoyerKasevich1997} P. Boyer and M. Kasevich, Phys. Rev. A \textbf{56}, R1083 (1997).
\bibitem{DunninghamBurnett2000} J. A. Dunningham and K. Burnett , Phys. Rev. A \textbf{61}, 065601 (2000).
\bibitem{DunBurnBarn2002} J. A. Dunningham, K. Burnett, and S. M. Barnett, Phys. Rev. Lett. \textbf{89}, 150401 (2002).
\bibitem{GerryCampos2003} C. C. Gerry and R. A. Campos, Phys. Rev. A \textbf{68}, 025602 (2003).
\bibitem{Kasevich2001} C. Orzel, A. K. Truchman, M. L. Fenselau,
and M. A. Kasevich, Science \textbf{291}, 2386 (2001).
\bibitem{Mott} M. Greiner, O. Mandel, T. Esslinger, T. W.
H{\"a}nsch, and I. Bloch, Nature \textbf{415}, 39 (2002).
\bibitem{Fisher} M. P. A. Fisher, P. B. Weichman, G. Grinstein,
and D. S. Fisher, Phys. Rev. B \textbf{40}, 546 (1989).
\bibitem{Zoller} D. Jaksch, C. Bruder, J. I. Cirac, C. W.
Gardiner, and P. Zoller, Phys. Rev. Lett. \textbf{81}, 3108 (1998).
\bibitem{Attractive} P. A. Ruprecht, M. J. Holland, K. Burnett,
and M. Edwards, Phys. Rev. A {\bf 51}, 4704 (1995); G. Baym, and
C. Pethick, Phys. Rev. Lett. {\bf 76}, 6 (1996); Y. Kagan, G. V.
Shlyapnikov, and J. T. M. Walraven, Phys. Rev. Lett. {\bf 76},
2670 (1996).
\bibitem{Steel} M. Steel and M. Collett, Phys. Rev. A \textbf{57}, 2920 (1998).
\bibitem{Cirac} J. I. Cirac, M. Lewenstein, K. M{\o}lmer, and P. Zoller, Phys. Rev. A \textbf{57}, 2920 (1998).
\bibitem{Ketterle} Y. Shin, M. Saba, T. A. Pasquini, W. Ketterle,
D. Pritchard, and A. E. Leanhardt, Phys. Rev. Lett. \textbf{92},
050405 (2004).
\bibitem{Milburn1997} G. J. Milburn, J. Corney, E. M. Wright, and
D. F. Walls, Phys. Rev. A \textbf{55}, 4318 (1997).
\bibitem{Weiping} J. F. Coburn, G. J. Milburn, and W. Zhang, Phys. Rev. A \textbf{59}, 4630 (1999).
\bibitem{Ueda1993} M. Kitagawa, and M. Ueda, Phys. Rev. A
\textbf{47}, 5138 (1993)
\bibitem{LuisSS} D. T. Pegg and S. M. Barnett, Phys. Rev A \textbf{39} 1665 (1989);  A. Luis and L. L. Sanchez-Soto, Phys. Rev. A \textbf{48}, 4702 (1993).
\bibitem{Javanainen} J. Javanainen and M. Yu, Ivanov, Phys. Rev. A \textbf{60}, 2351 (1999).
\bibitem{Kasevich2} M. Kasevich, private communication.
\end{thebibliography}
\end{document}